\input harvmac
\input epsf
%
\newbox\hdbox%
\newcount\hdrows%
\newcount\multispancount%
\newcount\ncase%
\newcount\ncols
\newcount\nrows%
\newcount\nspan%
\newcount\ntemp%
\newdimen\hdsize%
\newdimen\newhdsize%
\newdimen\parasize%
\newdimen\spreadwidth%
\newdimen\thicksize%
\newdimen\thinsize%
\newdimen\tablewidth%
\newif\ifcentertables%
\newif\ifendsize%
\newif\iffirstrow%
\newif\iftableinfo%
\newtoks\dbt%
\newtoks\hdtks%
\newtoks\savetks%
\newtoks\tableLETtokens%
\newtoks\tabletokens%
\newtoks\widthspec%
%
%
%
%
\tableinfotrue%
\catcode`\@=11
%
%
\def\tstrut{\vrule height3.1ex depth1.2ex width0pt}%
\def\and{\char`\&}
\def\tablerule{\noalign{\hrule height\thinsize depth0pt}}%
\thicksize=1.5pt
\thinsize=0.6pt
\def\thickrule{\noalign{\hrule height\thicksize depth0pt}}%
\def\ctr#1{\hfil\ #1\hfil}%
%
%
%
%
\tablewidth=-\maxdimen%
\spreadwidth=-\maxdimen%
\def\tabskipglue{0pt plus 1fil minus 1fil}%
%
%
\centertablestrue%
%
%
%
%
\parasize=4in%
\gdef\ARGS{########}
\gdef\headerARGS{####}
\def\@mpersand{&}
{\catcode`\|=13
\gdef\letbarzero{\let|0}
\gdef\letbartab{\def|{&&}}%
\gdef\letvbbar{\let\vb|}%
}
{\catcode`\&=4
\def\ampskip{&\omit\hfil&}
\catcode`\&=13
\let&0
\xdef\letampskip{\def&{\ampskip}}%
\gdef\letnovbamp{\let\novb&\let\tab&}
}
\def\begintable{
   \begingroup%
   \catcode`\|=13\letbartab\letvbbar%
   \catcode`\&=13\letampskip\letnovbamp%
   \def\multispan##1{
      \omit \mscount##1%
      \multiply\mscount\tw@\advance\mscount\m@ne%
      \loop\ifnum\mscount>\@ne \sp@n\repeat%
   }
   \def\|{%
      &\omit\widevline&%
   }%
   \ruledtable
}
\long\def\ruledtable#1\endtable{%
%
%
%
   \offinterlineskip
   \tabskip 0pt
   \def\widevline{\vrule width\thicksize}
   \def\endrow{\@mpersand\omit\hfil\crnorm\@mpersand}%
   \def\crthick{\@mpersand\crnorm\thickrule\@mpersand}%
   \def\crthickneg##1{\@mpersand\crnorm\thickrule
          \noalign{{\skip0=##1\vskip-\skip0}}\@mpersand}%
   \def\crnorule{\@mpersand\crnorm\@mpersand}%
   \def\crnoruleneg##1{\@mpersand\crnorm
          \noalign{{\skip0=##1\vskip-\skip0}}\@mpersand}%
   \let\nr=\crnorule
   \def\endtable{\@mpersand\crnorm\thickrule}%
   \let\crnorm=\cr
%
%
   \edef\cr{\@mpersand\crnorm\tablerule\@mpersand}%
   \def\crneg##1{\@mpersand\crnorm\tablerule
          \noalign{{\skip0=##1\vskip-\skip0}}\@mpersand}%
   \let\ctneg=\crthickneg
   \let\nrneg=\crnoruleneg
   \the\tableLETtokens
%
%
   \tabletokens={&#1}
%
%
   \countROWS\tabletokens\into\nrows%
   \countCOLS\tabletokens\into\ncols%
%
%
   \advance\ncols by -1%
   \divide\ncols by 2%
   \advance\nrows by 1%
%
%
   \iftableinfo %
      \immediate\write16{[Nrows=\the\nrows, Ncols=\the\ncols]}%
   \fi%
%
%
   \ifcentertables
      \ifhmode \par\fi
      \line{
      \hss
   \else %
      \hbox{%
   \fi
      \vbox{%
         \makePREAMBLE{\the\ncols}
         \edef\next{\preamble}
         \let\preamble=\next
         \makeTABLE{\preamble}{\tabletokens}
      }
      \ifcentertables \hss}\else }\fi
   \endgroup
   \tablewidth=-\maxdimen
   \spreadwidth=-\maxdimen
}
\def\makeTABLE#1#2{
   {
   \let\ifmath0
   \let\header0
   \let\multispan0
%
%
   \ncase=0%
   \ifdim\tablewidth>-\maxdimen \ncase=1\fi%
   \ifdim\spreadwidth>-\maxdimen \ncase=2\fi%
   \relax
%
   \ifcase\ncase %
      \widthspec={}%
   \or %
      \widthspec=\expandafter{\expandafter t\expandafter o%
                 \the\tablewidth}%
   \else %
      \widthspec=\expandafter{\expandafter s\expandafter p\expandafter r%
                 \expandafter e\expandafter a\expandafter d%
                 \the\spreadwidth}%
   \fi %
   \xdef\next{
      \halign\the\widthspec{%
      #1
      \noalign{\hrule height\thicksize depth0pt}
      \the#2\endtable
%
      }
   }
   }
   \next
}
\def\makePREAMBLE#1{
   \ncols=#1
   \begingroup
   \let\ARGS=0
   \edef\xtp{\widevline\ARGS\tabskip\tabskipglue%
   &\ctr{\ARGS}\tstrut}
   \advance\ncols by -1
   \loop
      \ifnum\ncols>0 %
      \advance\ncols by -1%
      \edef\xtp{\xtp&\vrule width\thinsize\ARGS&\ctr{\ARGS}}%
   \repeat
   \xdef\preamble{\xtp&\widevline\ARGS\tabskip0pt%
   \crnorm}
   \endgroup
}
\def\countROWS#1\into#2{
   \let\countREGISTER=#2%
   \countREGISTER=0%
   \expandafter\ROWcount\the#1\endcount%
}%
\def\ROWcount{%
   \afterassignment\subROWcount\let\next= %
}%
\def\subROWcount{%
   \ifx\next\endcount %
      \let\next=\relax%
   \else%
      \ncase=0%
      \ifx\next\cr %
         \global\advance\countREGISTER by 1%
         \ncase=0%
      \fi%
      \ifx\next\endrow %
         \global\advance\countREGISTER by 1%
         \ncase=0%
      \fi%
      \ifx\next\crthick %
         \global\advance\countREGISTER by 1%
         \ncase=0%
      \fi%
      \ifx\next\crnorule %
         \global\advance\countREGISTER by 1%
         \ncase=0%
      \fi%
      \ifx\next\crthickneg %
         \global\advance\countREGISTER by 1%
         \ncase=0%
      \fi%
      \ifx\next\crnoruleneg %
         \global\advance\countREGISTER by 1%
         \ncase=0%
      \fi%
      \ifx\next\crneg %
         \global\advance\countREGISTER by 1%
         \ncase=0%
      \fi%
      \ifx\next\header %
         \ncase=1%
      \fi%
      \relax%
      \ifcase\ncase %
         \let\next\ROWcount%
      \or %
         \let\next\argROWskip%
      \else %
      \fi%
   \fi%
   \next%
}
\def\counthdROWS#1\into#2{%
\dvr{10}%
   \let\countREGISTER=#2%
   \countREGISTER=0%
\dvr{11}%
\dvr{13}%
   \expandafter\hdROWcount\the#1\endcount%
\dvr{12}%
}%
\def\hdROWcount{%
   \afterassignment\subhdROWcount\let\next= %
}%
\def\subhdROWcount{%
   \ifx\next\endcount %
      \let\next=\relax%
   \else%
      \ncase=0%
      \ifx\next\cr %
         \global\advance\countREGISTER by 1%
         \ncase=0%
      \fi%
      \ifx\next\endrow %
         \global\advance\countREGISTER by 1%
         \ncase=0%
      \fi%
      \ifx\next\crthick %
         \global\advance\countREGISTER by 1%
         \ncase=0%
      \fi%
      \ifx\next\crnorule %
         \global\advance\countREGISTER by 1%
         \ncase=0%
      \fi%
      \ifx\next\header %
         \ncase=1%
      \fi%
\relax%
      \ifcase\ncase %
         \let\next\hdROWcount%
      \or%
         \let\next\arghdROWskip%
      \else %
      \fi%
   \fi%
   \next%
}%
{\catcode`\|=13\letbartab
\gdef\countCOLS#1\into#2{%
   \let\countREGISTER=#2%
   \global\countREGISTER=0%
   \global\multispancount=0%
   \global\firstrowtrue
   \expandafter\COLcount\the#1\endcount%
   \global\advance\countREGISTER by 3%
   \global\advance\countREGISTER by -\multispancount
}%
\gdef\COLcount{%
   \afterassignment\subCOLcount\let\next= %
}%
{\catcode`\&=13%
\gdef\subCOLcount{%
   \ifx\next\endcount %
      \let\next=\relax%
   \else%
      \ncase=0%
      \iffirstrow
         \ifx\next& %
            \global\advance\countREGISTER by 2%
            \ncase=0%
         \fi%
         \ifx\next\span %
            \global\advance\countREGISTER by 1%
            \ncase=0%
         \fi%
         \ifx\next| %
            \global\advance\countREGISTER by 2%
            \ncase=0%
         \fi
         \ifx\next\|
            \global\advance\countREGISTER by 2%
            \ncase=0%
         \fi
         \ifx\next\multispan
            \ncase=1%
            \global\advance\multispancount by 1%
         \fi
         \ifx\next\header
            \ncase=2%
         \fi
         \ifx\next\cr       \global\firstrowfalse \fi
         \ifx\next\endrow   \global\firstrowfalse \fi
         \ifx\next\crthick  \global\firstrowfalse \fi
         \ifx\next\crnorule \global\firstrowfalse \fi
         \ifx\next\crnoruleneg \global\firstrowfalse \fi
         \ifx\next\crthickneg  \global\firstrowfalse \fi
         \ifx\next\crneg       \global\firstrowfalse \fi
      \fi
\relax
      \ifcase\ncase %
         \let\next\COLcount%
      \or %
         \let\next\spancount%
      \or %
         \let\next\argCOLskip%
      \else %
      \fi %
   \fi%
   \next%
}%
\gdef\argROWskip#1{%
   \let\next\ROWcount \next%
}
\gdef\arghdROWskip#1{%
   \let\next\ROWcount \next%
}
\gdef\argCOLskip#1{%
   \let\next\COLcount \next%
}
}
}
\def\spancount#1{
   \nspan=#1\multiply\nspan by 2\advance\nspan by -1%
   \global\advance \countREGISTER by \nspan
   \let\next\COLcount \next}%
\def\dvr#1{\relax}%
\def\header#1{%
\dvr{1}{\let\cr=\@mpersand%
\hdtks={#1}%
\counthdROWS\hdtks\into\hdrows%
\advance\hdrows by 1%
\ifnum\hdrows=0 \hdrows=1 \fi%
\dvr{5}\makehdPREAMBLE{\the\hdrows}%
\dvr{6}\getHDdimen{#1}%
{\parindent=0pt\hsize=\hdsize{\let\ifmath0%
\xdef\next{\valign{\headerpreamble #1\crnorm}}}\dvr{7}\next\dvr{8}%
}%
}\dvr{2}}
\def\makehdPREAMBLE#1{
\dvr{3}%
\hdrows=#1
{
\let\headerARGS=0%
\let\cr=\crnorm%
\edef\xtp{\vfil\hfil\hbox{\headerARGS}\hfil\vfil}%
\advance\hdrows by -1
\loop
\ifnum\hdrows>0%
\advance\hdrows by -1%
\edef\xtp{\xtp&\vfil\hfil\hbox{\headerARGS}\hfil\vfil}%
\repeat%
\xdef\headerpreamble{\xtp\crcr}%
}
\dvr{4}}
\def\getHDdimen#1{%
\hdsize=0pt%
\getsize#1\cr\end\cr%
}
\def\getsize#1\cr{%
\endsizefalse\savetks={#1}%
\expandafter\lookend\the\savetks\cr%
\relax \ifendsize \let\next\relax \else%
\setbox\hdbox=\hbox{#1}\newhdsize=1.0\wd\hdbox%
\ifdim\newhdsize>\hdsize \hdsize=\newhdsize \fi%
\let\next\getsize \fi%
\next%
}%
\def\lookend{\afterassignment\sublookend\let\looknext= }%
\def\sublookend{\relax%
\ifx\looknext\cr %
\let\looknext\relax \else %
   \relax
   \ifx\looknext\end \global\endsizetrue \fi%
   \let\looknext=\lookend%
    \fi \looknext%
}%
%
%
\def\tablelet#1{%
   \tableLETtokens=\expandafter{\the\tableLETtokens #1}%
}%
\catcode`\@=12
\def \inparg{\leftskip = 40 pt\rightskip = 40pt}
\def \outparg{\leftskip = 0 pt\rightskip = 0pt}

\thicksize=0.7pt 
\thinsize=0.5pt
\def\ctr#1{\hfil $\,\,\,#1\,\,\,$ \hfil}
\def\tstrut{\vrule height 2.7ex depth 1.0ex width 0pt}

\def\frak#1#2{{\textstyle{{#1}\over{#2}}}}
\def\frakk#1#2{{{#1}\over{#2}}}
\def\mbar{{\overline{m}}}
\def\mhat{{\hat m }}
\def\semi{;\hfil\break}
\def\npb{{Nucl.\ Phys.\ }{\bf B}}
\def\prd{{Phys.\ Rev.\ }{\bf D}}
\def\prl{Phys.\ Rev.\ Lett.\ }
\def\plb{{Phys.\ Lett.\ }{\bf B}}
\def\lf{16\pi^2}

\def\Qbar{{\overline Q}}

\def\ttil{\tilde t}
\def\btil{\tilde b}
\def\tautil{\tilde \tau}
\def\util{\tilde u}
\def\dtil{\tilde d}
\def\etil{\tilde e}

\def\nutil{\tilde \nu} 
\def\chitil{\tilde \chi} 
\def\TeV{{\rm TeV}}
\def\GeV{{\rm GeV}}
{\nopagenumbers
\line{\hfil LTH 477}
\line{\hfil hep-ph/0003081}
\vskip .5in
\centerline{\titlefont Fayet-Iliopoulos $D$-terms and } 
\centerline{\titlefont anomaly mediated supersymmetry breaking}
\vskip 1in
\centerline{\bf I.~Jack and D.R.T.~Jones}
\bigskip
\centerline{\it Dept. of Mathematical Sciences,
University of Liverpool, Liverpool L69 3BX, U.K.}
\vskip .3in
We show that in a minimal extension of the MSSM by means of an extra 
$U_1$ gauge group, the negative mass-squared problem characteristic of
the  Anomaly Mediated Supersymmetry Breaking scenario is naturally 
solved by means of Fayet-Iliopoulos $D$-terms. We derive a set of sum
rules  for the sparticle  masses which are consequences of the resulting
framework.

\Date{March 2000}}


The MSSM has (according to a recent census\ref\haber{H.E.~Haber,  Nucl.
Phys. B (Proc. Suppl.) 62A-C (1998) 469}) 124 parameters;  an  obvious
embarrassment, and any (principled)  reduction of this alarming total is
obviously worthy of examination.  Hence there has been interest in a 
specific and  predictive  framework wherein the 
gaugino masses $M_a$, the $\phi^3$ coupling
$h^{ijk}$ and the  $\phi\phi^*$-mass $m^i{}_j$ are all given in terms  of
a single mass parameter, $m_0$,  and the $\beta$-functions of the
unbroken theory  by simple relations that are renormalisation group
(RG) invariant. These results for the 
soft terms were (with the exception of the solution for the gaugino mass) 
first developed by seeking solutions to the exact $\beta$-function equations
\ref\jjpa{I.~Jack, D.R.T.~Jones and A.~Pickering,
\plb426 (1998) 73}\ref\akk{L.V.~Avdeev, D.I.~Kazakov and
I.N.~Kondrashuk, \npb510 (1998) 289}; remarkably, it was then shown 
\ref\lisa{L. Randall and R. Sundrum, \npb 557 (1999) 79}
\ref\glmr{G.F. Giudice, M.A. Luty, H. Murayama and  R. Rattazzi,
JHEP 9812 (1998) 27}
that  they arise naturally if the supersymmetry-breaking terms originate in a
vacuum expectation value for an auxiliary field in the  supergravity 
multiplet. 
In this scenario, termed  `Anomaly Mediated Supersymmetry Breaking' 
(AMSB), $m_0$ is in fact the gravitino mass, and   
all the gaugino masses,  soft $\phi\phi^*$ masses and
$A$-parameters are determined in terms of it
\ref\appp{A. Pomarol and  R. Rattazzi, JHEP 9905 (1999) 013}%
\nref\ggw{T. Gherghetta, G.F. Giudice and J.D. Wells, \npb 559 (1999) 27}%
\nref\clmp{Z. Chacko, M.A. Luty, I. Maksymyk and E. Ponton, hep-ph/9905390}%
\nref\kss{E. Katz, Y. Shadmi and Y. Shirman, JHEP 9908 (1999) 015}%
\nref\jlftm{J.L.~Feng and T.~Moroi, hep-ph/9907319}%
\nref\gdk{G.D.~Kribs, hep-ph/9909376}%
\nref\shusu{S.~Su, hep-ph/9910481}%
\nref\rzzsw{R. Rattazzi, A. Strumia and J.D. Wells, hep-ph/9912390}%
--\ref\fepjw{F.E.~Paige and J. Wells, hep-ph/0001249}.  
Unfortunately, however, a minimal implementation 
leads inevitably to negative $(\hbox{mass})^2$ sleptons. 
The simplest resolution is the introduction of  a common scalar 
$(\hbox{mass})^2$, presumed to result from some other source of
supersymmetry breaking. The advantage of this is that only one new parameter is 
introduced: the disadvantage is that RG-invariance of the soft mass 
prediction is sacrificed.    

Here we propose an alternative solution in which the extra source of
supersymmetry breaking arises spontaneously within the  low energy
effective field theory, by exploiting the fact that supersymmetric theories
including $U_1$ factors have (in general)  Fayet-Iliopoulos (FI)
$D$-terms.  In the MSSM, there is a non-zero  FI-term, but this cannot
solve the slepton problem because  its $(\hbox{mass})^2$ contributions
to the  LH and RH sleptons  have opposite signs,  being dictated by the
hypercharge of the relevant field. Our proposed solution involves
extending the MSSM to  incorporate an extra $U_1$. It then becomes 
possible for both LH and RH sleptons to
achieve the nirvana of positive $(\hbox{mass})^2$ via FI contributions
\foot{Use of FI
terms is  also a feature of Ref.~\kss, but in a different manner to that
proposed  here.}. 

Theories with an extra $U_1$ have been studied as a means of 
parameterising deviations from the SM, 
and also for more positive  reasons\foot{We note the
suggestion\ref\lang{J.~Erler and P.~Langacker, \prl 84 (2000) 212}\  that
there are already ``hints'' of the existence  of an extra $Z'$ at around
$1\TeV$.}. For example, in the
supersymmetric case an extra $U_1$ can be used to  explain the absence of
dimension-4 R-parity violation 
(operators violating baryon and lepton number)
\ref\weinb{S.~Weinberg, \prd 26 (1982) 287}%
\nref\fost{A.~Font, L.E. Ib\'a\~nez and F. Quevedo, 
\plb 228 (1989) 79}%
\nref\cham{A.H. Chamseddine and  H. Dreiner, \npb 447 (1995) 195}%
--\ref\cheng{H-C.~Cheng, B.A.~Dobrescu and K.T.~Matchev, \plb439 (1998) 301; 
\npb543 (1999) 47}.
Here  we consider a minimal anomaly-free 
generalisation of the MSSM to the group ${\cal G}\otimes U'_1$, where 
${\cal G} = SU_3\otimes SU_2\otimes U_1$, with 
the addition of an unspecified number $N$ of
${\cal G}$ singlets ($S_i$) and a superpotential $W$ of the full 
theory  given by
\eqn\spt{
W = W_{MSSM} + W_S (S_i).}
Here 
\eqn\wmssm{W_{MSSM} = \mu_{s} H_1 H_2 + \lambda_t H_2
Q t^c + \lambda_b H_1 Q b^c  + \lambda_{\tau}H_1 L \tau^c.}
We retain Yukawa couplings only for the third generation,  
$Q,L,t^c,b^c,\tau^c$, and we will denote the corresponding fields 
of the other generations by $\Qbar, E, u^c, d^c, e^c$. 
Let us define the $U'_1$ hypercharges of the MSSM fields 
$Q,L,t^c,b^c,\tau^c,H_1,H_2$ to be 
$Y'_Q,Y'_L,Y'_{t^c},Y'_{b^c},Y'_{\tau^c}, Y'_{H_1},Y'_{H_2}$. 
We will assume that the quark and lepton assignments are generation 
independent, i.e. $Y'_L = Y'_E$ etc; this means that our model will, 
in fact, naturally suppress dangerous flavour violating processes.   
It is straightforward to show that gauge invariance and 
absence of mixed gauge  
anomalies involving  $U'_1$ leads to the relations\cham:
\eqn\anrels{\eqalign{Y'_{H_1} + Y'_L + Y'_{\tau^c} &= 
Y'_{H_2} -  Y'_L - Y'_{\tau^c} = 3Y'_Q + Y'_L = 0,\cr
3Y'_{t^c} + 2Y'_L + 3Y'_{\tau^c} &= 3Y'_{b^c}-4Y'_L-3Y'_{\tau^c} = 0.\cr}}
(To obtain these relations it is not necessary to assume 
that $Y'_{H_1} + Y'_{H_2}=0$: that is, gauge invariance of the $\mu_{s}$-term 
is a consequence of the framework\cham.) 
To cancel the $(U_1')^3$ and $U'_1$-gravitational anomalies, the 
$U_1'$ hypercharges $s_i$ of the fields $S_i$ must 
satisfy the constraints:
\eqn\sanom{
\sum_{i=1}^N s_i = -3(2Y'_L + Y'_{\tau^c}),\quad\hbox{and}\quad
\sum_{i=1}^N s_i^3 = -3(2Y'_L + Y'_{\tau^c})^3.}
Suppose we prefer hypercharges to be rational; then the 
classification of solutions to Eq.~\sanom\ is an example of 
a well-known problem: finding the 
rational points on a $n$-dimensional surface. For example the rational 
points on the circle $x^2 + y^2 =1$ are given by 
\eqn\circ{
(x,y) = (0,-1) \quad\hbox{and}\quad 
\left(\frakk{2q}{1+q^2},\frakk{1-q^2}{1+q^2}\right)}
where $q$ is rational.
The case $N=3$ of Eq.~\sanom\ was analysed in Ref.~\cham; the solution is 
\eqn\chameq{\eqalign{
(s_1,s_2,s_3) &= -(2Y'_L + Y'_{\tau^c})(1,1,1)\quad\hbox{and}\cr
(s_1,s_2,s_3) &= -(2Y'_L + Y'_{\tau^c})\left(\frakk{5+3q^2}{q^2-1},
\frakk{q^2+q+4}{q+1},-\frakk{q^2-q+4}{q-1}\right) \cr}}
where again $q$ is rational. 
We will simply assume that for some $N$ there exists an appropriate 
solution, and that the singlet sector provides the $Z'$ vector boson with 
a sufficiently large mass term so that its mixing with the $Z$ is adequately 
suppressed.

For simplicity we  also choose to impose the condition $\Tr (YY') =0$.
This  prevents mixing of the $U_1$ and $U'_1$ kinetic terms for the 
gauge bosons (through
the one loop approximation)\foot{The consequences of this kinetic 
mixing have
been studied  in Ref~\ref\kmr{
K.S.~Babu, C. Kolda and  J.~March-Russell, \prd 57 (1998) 6788\semi
K.R. Dienes, C. Kolda and  J.~March-Russell, \npb 492 (1997) 104}.} 
and leads to the relation:
\eqn\yyprime{
3Y'_L + 7Y'_{\tau^c} = 0}
The resulting hypercharges are shown in Table~1, with 
the $U_1$ ones for comparison:

\vskip3em
\vbox{
\begintable
| Q      | L           | t^c     | b^c         
     | \tau^c      | H_1         | H_2     | S_i\cr
Y    | \frakk{1}{6}|-\frakk{1}{2}| -\frakk{2}{3}
| \frakk{1}{3}|1    |-\frakk{1}{2}|\frakk{1}{2}| 0\cr
Y'   | \frakk{7}{3}|-7           | \frakk{5}{3} 
| -\frakk{19}{3}| 3 | 4           |-4          | s_i 
\endtable
\bigskip
\inparg
{\noindent {\it Table~1:\/} The $U_1$ and $U_1{}'$ hypercharges.}
\bigskip \outparg}
{\noindent With this assignment we indeed prevent the  dimension-4 R-parity 
violating  operators.}  

In a theory with anomaly-generated soft parameters,
and with FI terms $\xi_1 D$, $\xi_2D'$ for $U_1$ and $U_1'$ respectively, 
a soft mass for a generic field is given 
after 
elimination of the $D$-terms by 
$m^2+g'\xi_1Y+g^{\prime\prime}\xi_2Y',$ where  
\eqn\anoma{m^2=\frak{1}{2}|m_0|^2\mu\frakk{d}{d\mu}\gamma,}
with $\gamma$ being the anomalous dimension.  
(We denote the gauge couplings for $SU(3)$, $SU(2)$, 
the MSSM $U_1$ and the new $U_1'$ by $g_3$, $g_2$, $g_1=\sqrt{\frak{5}{3}}g'$, 
and $g^{\prime\prime}$ respectively.)    
Consequently, after spontaneous symmetry 
breaking, the effective soft masses of the squarks and sleptons
(before including $A$-parameter and $\mu_{s}$-term mixing effects) are given by
\eqn\mssrel{\eqalign{
\mbar^2_Q &= m^2_Q +\frak{1}{6}\zeta_1 +\zeta_2 Y'_Q,\quad
\mbar^2_{t^c} = m^2_{t^c} -\frak{2}{3}\zeta_1 + \zeta_2 Y'_{t^c},\cr
\mbar^2_{b^c} &= m^2_{b^c} +\frak{1}{3}\zeta_1  +\zeta_2 Y'_{b^c},\quad
\mbar^2_L =  m^2_L -\frak{1}{2}\zeta_1 + \zeta_2 Y'_L,\cr
\mbar^2_{\tau^c} &= m^2_{\tau^c} +\zeta_1 + \zeta_2 Y'_{\tau^c},\cr}}
where 
\eqn\xidefs{\zeta_1 = g'[\xi_1 - g'\frak{1}{4}(v_1^2-v_2^2)], \quad
\zeta_2 = g^{\prime\prime}[\xi_2 + {\cal S} + 2g^{\prime\prime}(v_1^2-v_2^2)],}
and where 
\eqn\anomb{
m_Q^2=\frak{1}{2}|m_0|^2\mu\frakk{d}{d\mu}\gamma_Q, \quad 
m_{t^c}^2=\frak{1}{2}|m_0|^2\mu\frakk{d}{d\mu}\gamma_{t^c},}
and so on. It is easy to write down the analogous expressions for the 
other generations. We have included in Eqs.~\mssrel,\xidefs\ the standard 
$D$-term contributions to the masses resulting from the Higgs vevs, together
with a contribution ${\cal S}$ from the (unknown) vevs of the singlets
$S_i$. Note that the dependence
on the singlet sector is subsumed into $\zeta_2$, and 
therefore much of the discussion can be independent of the precise structure of 
the singlet terms.

The relation between each $\mbar^2$ and $m^2$ in  Eq.~\mssrel\ is 
quite generally RG invariant (it is important that 
the $\beta_{\mbar^2},\beta_{m^2}$ are calculated with 
$D$ eliminated and $D$ uneliminated respectively
\ref\jjfi{I.~Jack and D.R.T. Jones, \plb 473 (2000) 102}). 
It is
also invariant if we replace $\zeta_{1,2}$ by constants (but in this case 
both $\beta_{\mbar^2}$ and $\beta_{m^2}$ are calculated with
$D$ uneliminated). Thus in 
a general theory with ${\cal N}$ non-anomalous 
 $U_1$ factors, then the relation
\eqn\notinvar{
(\mhat^2)^i{}_j = (m^2)^i{}_j + m_0^2\delta^i{}_j}
is not RG invariant (for constant $m_0^2$), but 
\eqn\invar{
(\mhat^2)^i{}_j = (m^2)^i{}_j + m_0^2\sum_{a=1}^{\cal N} k_a (Y_a)^i{}_j}
{\it is} RG invariant.   
This is easily shown using the gauge invariance and  anomaly
cancellation conditions, together with the general formula  for
$\beta_{m^2}$ given, for example in Ref.~\jjfi. Evidently this
invariance  continues to hold in the limit that the $U_1$ gauge
couplings approach zero,  so we do not even need the $U_1$ groups
to be gauged (or to impose relations like Eq.~\yyprime, so that 
we could then have the same sign for $Y'_L$ and $Y'_{\tau^c}$); 
though clearly it would be artificial to impose
anomaly cancellation conditions in that case. String theories often give rise 
to apparently {\it anomalous\/} $U'_1$ symmetries, with the anomaly 
cancelled by the Green-Schwarz mechanism. We might therefore 
entertain the possibility of dispensing with the singlet sector
and invoking the GS mechanism to cancel the $(U'_1)^3$ 
and $U'_1$-gravitational anomalies (see Eq.~\sanom). If the $U_1'$ symmetry
were broken at a high mass scale, the only low-energy residue of the 
$U_1'$ would be the FI terms.   
However, we would then lack a 
rationale for imposing cancellation of the mixed gauge anomalies, 
a cancellation necessary to make Eq.~\invar\ RG invariant. We 
will therefore persist with a gauged, non-anomalous $U'_1$.   
 
The gaugino mass for a gauge coupling $g$ (either $g_3$, $g_2$, $g_1$ 
or $g^{\prime\prime}$)
in the AMSB scenario is given by
\foot{The significance of the {\it sign\/} of the gluino mass term 
is investigated in Ref.~\gdk.}
\eqn\anomd{
M_g=m_0{\beta_g\over{g}}.}
Moreover, the $A$-parameters are given by
\eqn\anomc{
A_t=-m_0(\gamma_Q+\gamma_{t^c}+\gamma_{H_2}), \quad 
A_b=-m_0(\gamma_Q+\gamma_{b^c}+\gamma_{H_1}), \quad
A_{\tau}=-m_0(\gamma_L+\gamma_{\tau^c}+\gamma_{H_1}).}
(We could write down similar results for the first two
generation $A$ parameters, but they will have no impact on our calculations
since the corresponding Yukawa couplings are small.)

For completeness we record here the expressions for the anomalous dimensions:
\eqn\gams{\eqalign{
\lf\gamma_{H_1} &= 3\lambda_b^2+\lambda_{\tau}^2-\frak32g_2^2
-\frak{3}{10}g_1^2
-2Y_{H_1}^{\prime2}g^{\prime\prime2},\cr
\lf\gamma_{H_2} &= 3\lambda_t^2-\frak32g_2^2-\frak{3}{10}g_1^2
-2Y_{H_2}^{\prime2}g^{\prime\prime2},\cr
\lf\gamma_{L} &= \lambda_{\tau}^2-\frak32g_2^2-\frak{3}{10}g_1^2
-2Y_L^{\prime2}g^{\prime\prime2},\cr
\lf\gamma_{Q} &= \lambda_b^2+\lambda_t^2-\frak83g_3^2-\frak32g_2^2
-\frak{1}{30}g_1^2
-2Y_Q^{\prime2}g^{\prime\prime2},\cr
\lf\gamma_{t^c} &= 2\lambda_t^2-\frak83g_3^2-\frak{8}{15}g_1^2
-2Y_{t^c}^{\prime2}g^{\prime\prime2},\cr
\lf\gamma_{b^c} &= 2\lambda_b^2-\frak83g_3^2-\frak{2}{15}g_1^2
-2Y_{b^c}^{\prime2}g^{\prime\prime2},\cr
\lf\gamma_{\tau^c} &= 2\lambda_{\tau}^2-\frak65g_1^2
-2Y_{\tau^c}^{\prime2}g^{\prime\prime2}.\cr}}
In the tree approximation the $\mu_{s}$-term is given by 
the Higgs minimisation condition:
\eqn\muterm{
\mu_{s}^2 = {m^2_{H_1}-m^2_{H_2} \tan^2\beta\over{\tan^2\beta-1}}
-\frakk{1}{2}M_W^2 
+ \sec 2\beta \left(\frakk{1}{2}\zeta_1 -4 \zeta_2\right).}
The masses of the pseudoscalar and charged Higgs bosons are given at 
leading order by the usual expressions
\eqn\higgsa{m_A^2 = 2r_1, \quad m_{H^{\pm}}^2=2r_1+M_W^2}
where we define 
\eqn\rterms{
r_1 = \frak{1}{2}(m^2_{H_2}+m^2_{H_1})+ \mu_{s}^2.}
The other minimisation condition,
\eqn\mthree{
m_3^2 = r_1\sin 2\beta}
determines the soft $H_1H_2$ mass term. 
In fact there exists an RG invariant 
solution for this as well\jjpa:
\eqn\mthreetraj{
m_3^2 = -m_0\mu\frakk{d}{d\mu}\mu_s.}
We find, however, that there is no value of $m_0$ leading to 
an otherwise acceptable spectrum and a result for $\tan\beta$
satisfying Eq.~\mthree. Thus, in common with 
previous work on the AMSB scenario, we are obliged to assume that 
$m_3^2$ arises from an alternative source of supersymmetry breaking, 
presumably linked to the $\mu_{s}$-term. It is also possible to 
construct (perturbatively) a  RG 
trajectory for $\xi_{1,2}$ so that $\xi_{1,2} \sim m_0^2$
\jjfi, but the 
resulting values of $\zeta_{1,2}$ are too small for our purpose here. 

We choose to normalise  the $U'_1$ hypercharge so as to
satisfy at the weak scale the relation  
\eqn\ynorm{ \Tr (Y^2 g_1^2) =
\Tr(Y^{\prime 2}g^{\prime\prime 2}),} 
which corresponds to equal $U_1$ and $U_1'$ gaugino masses. 
We will present results for the
case when the  $\sum s_i^2$ is large, so that the $U'_1$ gauge coupling 
is small; this limit suppresses $Z-Z'$ mixing, while allowing a large 
$Z'$ mass (because $\sum s_i^2$ is large); though of course in this limit 
the $Z'$ would decouple in any case.  

Let us now consider the nature of the predicted mass spectrum. 
The heaviest sparticle masses scale with $m_0$ and are given roughly 
by $M_{\rm SUSY} = \frak{1}{40}m_0$. Consequently we 
take account of leading-log corrections by evaluating the mass 
spectrum at the scale $M_{\rm SUSY}$. 
In other words, before applying Eqs.~\mssrel, \muterm\ etc.,
we evolve the dimensionless couplings (together with $v_1$, $v_2$)
from the weak scale up to the scale $M_{\rm SUSY}$. 
In order that the sleptons have positive $(\hbox{mass})^2$, we require 
\eqn\triang{m_E^2-\frak{1}{2}\zeta_1 + \zeta_2 Y'_L > 0, \quad\hbox{and}\quad
m_{e^c}^2+\zeta_1 + \zeta_2 Y'_{\tau^c}>0,} 
where $m_E^2$ and $m_{e^c}^2$ are the standard AMSB expressions as in 
Eq.~\anomb.
It turns out that the most important other constraint comes from 
requiring $m_A^2 > 0$. This constraint, together with Eq.~\triang, define 
a triangular region in the $\zeta_1,\zeta_2$ plane. For $m_0=40\hbox{TeV}$, 
and for $\tan\beta=5$, this triangular region is shown in Fig.~1. 
\smallskip
\epsfysize= 2.5in
\centerline{\epsfbox{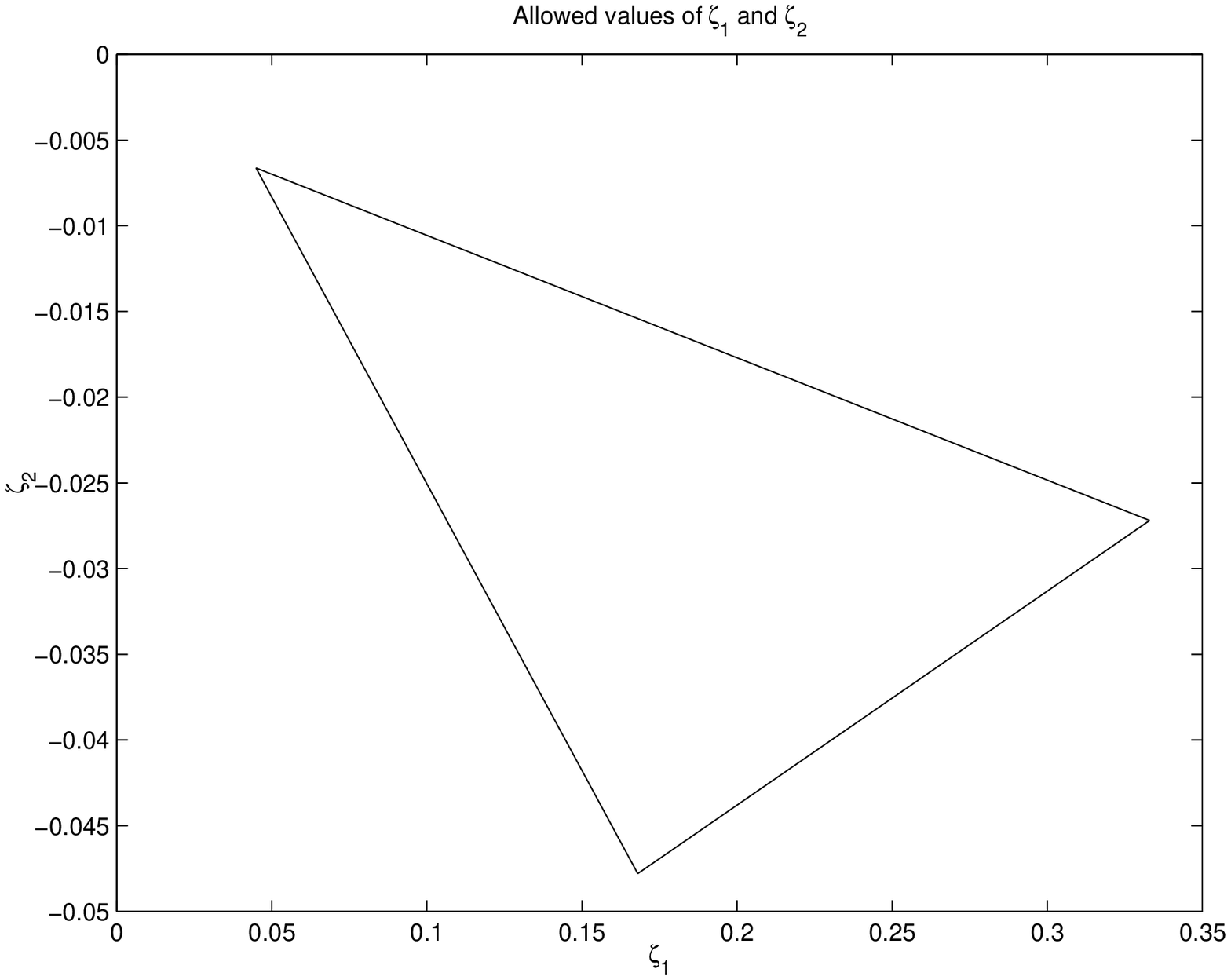}}
\inparg
{\it \noindent Fig.~1:\/}  
Allowed values of $\zeta_1$ and $\zeta_2$, for $\tan \beta=5$ and
$m_0=40\TeV$. 
\medskip
\outparg
For a choice of $\zeta_1=0.2$,
 $\zeta_2=-.02$, we find $|\mu_s| = 645\GeV$ and (choosing $\mu_s > 0$) a 
mass spectrum given by:
\eqn\spect{\eqalign{
m_{\ttil_1}&=575, \quad m_{\ttil_2}=861, \quad m_{\btil_1}=825, \quad 
m_{\btil_2}=1040, 
\quad m_{\tautil_1}=137,
\quad m_{\tautil_2}=339,\cr
m_{\util_L}&=931, \quad m_{\util_R}=851,\quad m_{\dtil_L}=935,
\quad m_{\dtil_R}=1045,
\quad m_{\etil_L}=139,\quad m_{\etil_R}=339,\cr
m_{\nutil} &= 112,\quad m_A=453, \quad m_{H^{\pm}} = 461, 
\quad m_{\tilde\chi_{1,2}^{\pm}} = 104, 649\quad 
m_{\tilde g} = 1007,}}
where all masses are given in GeV. 
The sleptons $\tautil_1$ and $\etil_L$ are light 
because we have chosen a point relatively near one edge. 
Alternative choices of $\zeta_{1,2}$ in the interior of the
allowed triangle lead
to a generally similar spectrum; well away from the edges 
$m_{\tautil_1}$ and $m_{\etil_L}$ approach $300\GeV$. 
The CP-even Higgs and neutralino
masses are  sensitive to the singlet sector so we cannot specify them
precisely. 
However based on the arguments of, for example, 
Ref.~\ref\kane{G.L. Kane, C. Kolda and  J.D. Wells, 
\prl 70 (1993) 2686}\ 
there will be an upper bound on the lighter Higgs of around $140\GeV$. 
Because $M_2$ is the smallest gaugino mass, 
we also expect a light neutralino approximately degenerate with 
the light chargino (both being predominantly wino in content)  
at around $104\GeV$, with the chargino being heavier due to 
radiative corrections. 
The light neutralino may be the LSP; 
the resulting distinctive phenomenology and the characteristic decay 
$\chitil^{\pm} \to \chitil^0 + \pi^{\pm}$ are described in 
Refs.~\ggw, \fepjw, \ref\fmrs{J.L.~Feng et al, \prl 83 (1999) 1731},
\ref\gumr{ J.F. Gunion and S. Mrenna , hep-ph/9906270}. 

In a limit such that the  singlet sector decouples, the CP even 
and neutralino spectrum become calculable and we obtain 
(for the same values of $\zeta_{1,2}$)
\eqn\hggseven{m_{h,H} = 88, 455\GeV}
and 
\eqn\neut{
m_{\chitil_{1\cdots4}} = 103, 366, 648, 658\GeV.}
As usual a complete calculation of the radiative corrections 
to $m_h$ may be expected to result in a somewhat higher value.
In this scenario there is no motivation for imposing Eq.~\yyprime;
but choosing a different set of $Y'$ satisfying Eq.~\anrels\ simply amounts
to a different choice of co-ordinates for the $(\zeta_1,\zeta_2)$ plane.

If $\zeta_{1,2}$ were to correspond to a point near one of the two 
appropriate edges of the triangle, the LSP would be a charged scalar lepton. 
Of course 
anomalous heavy isotope searches 
suggest that that a charged LSP is unlikely, but 
for a contrarian viewpoint on this issue, see for example  
Ref.~\ref\gouv{
A. de Gouv\^ea, A. Friedland and H. Murayama, \prd 59 (1999) 095008},
which is also of interest in that it 
considers the phenomenological footprints of  a FI term in the
MSSM.  

As previous authors have observed\ggw, $m_E^2$ and $m_{e^c}^2$ 
are very nearly equal; this does not extend to 
the physical masses $m_{\etil_L}$ and 
$m_{\etil_R}$ in  our framework, because of the FI contributions (the same 
observation applies to some other resolutions of the tachyonic 
slepton problem, see e.g. Ref.~\appp). 
Finally, the lightest strongly-interacting particle is the lighter stop, 
$\ttil_1$; but this is a feature of much of MSSM parameter space. 

As we reduce $m_0$, or increase
$\tan\beta$, the triangular region of $\zeta_{1,2}$ satisfying
Eq.~\triang\ and $m_A^2>0$ diminishes, and moreover, experimental
constraints on $m_{\tilde\chi^{\pm}_1}$ or $m_{\tautil_1}$ further
reduce the allowed region for smaller $m_0$ or large $\tan\beta$
respectively. In fact, we find  that an acceptable spectrum is only
possible for  $m_0\ge35\TeV$ (with $\tan\beta=5$) or for
$\tan\beta\le27$ (with $m_0=40\TeV$). For smaller $\tan\beta$, 
the spectrum is similar to Eq.~\spect, 
but the allowed triangle begins to shrink 
as $\tan\beta\to 2$, a value approaching (as it happens) the 
quasi-infra-red fixed point for $\lambda_t$.

We have taken $g^{\prime\prime}$ very small by taking $\sum s_i^2$ large and
imposing Eq.~\ynorm. For larger values of $g^{\prime\prime}$ the allowed 
parameter space is still determined by the triangle, and the broad 
features of the spectrum remain the same. 

The most distinctive feature of the model presented here is the existence 
of sum rules for combinations of masses in which the dependence on  
$\zeta_{1,2}$ cancels. We find
\eqn\sumrule{\eqalign{
\mbar_L^2+3\mbar_Q^2 &= m_L^2+3m_{Q}^2,\cr
\mbar_{t^c}^2+\mbar_{b^c}^2+2\mbar_Q^2 &= m_{t^c}^2+m_{b^c}^2+2m_Q^2,\cr
\mbar_{t^c}^2+\mbar_{\tau^c}^2-2\mbar_Q^2 &= m_{t^c}^2+m_{\tau^c}^2-2m_Q^2,\cr}}
where $\mbar^2$ are the effective soft mass parameters and $m^2$ are the 
pure AMSB masses as given by Eq.~\anomb. From these results we can obtain the 
following relations for the physical masses: 
\eqn\sumthr{\eqalign{
m_{\ttil_1}^2+m_{\ttil_2}^2+m_{\btil_1}^2+m_{\btil_2}^2 
- 2(m_t^2 + m_b^2)&= 
2.79\left(\frakk{m_0}{40}\right)^2\TeV^2 \cr
m_{\tautil_1}^2+m_{\tautil_2}^2+m_{\ttil_1}^2+m_{\ttil_2}^2 
- 2(m_t^2 + m_{\tau}^2) &= 1.15\left(\frakk{m_0}{40}\right)^2\TeV^2.\cr}}  

Similar results
apply for the first two generations as follows:
\eqn\sumrulea{\eqalign{
m_{\etil_L}^2+2m_{\util_L}^2+m_{\dtil_L}^2 &= m_E^2+3m_{\Qbar}^2 
= 2.63\left(\frakk{m_0}{40}\right)^2\TeV^2,\cr
m_{\util_R}^2+m_{\dtil_R}^2+m_{\util_L}^2+m_{\dtil_L}^2 &= 
m_{u^c}^2+m_{d^c}^2+2m_{\Qbar}^2 = 3.56\left(\frakk{m_0}{40}\right)^2\TeV^2,\cr
m_{\util_L}^2+m_{\dtil_L}^2-m_{\util_R}^2-m_{\etil_R}^2 &= 
2m_{\Qbar}^2  - m_{u^c}^2-m_{e^c}^2 = 
0.90\left(\frakk{m_0}{40}\right)^2\TeV^2.\cr}}
Note that in Eqs.~\sumthr, \sumrulea\ the dependence of the physical masses 
on $M_W^2$ has cancelled in the combinations on the 
left-hand side, in addition to the dependence on $\zeta_{1,2}$. 
Finally, sum rules involving the $CP$-odd Higgs:
\eqn\masumr{
m_A^2 - 2\sec 2\beta\left (m_{\etil_L}^2 +m_{\etil_R}^2\right) =
\sec 2\beta\left[m^2_{H_2} - m^2_{H_1} - 2(m^2_{e^c} + m^2_E)\right]
=0.49\left(\frakk{m_0}{40}\right)^2\TeV^2,}
and
\eqn\masumrb{\eqalign{
m_A^2 - 2\sec 2\beta\left (m_{\tautil_1}^2 + m_{\tautil_2}^2
-2m_{\tau}^2\right) &=
\sec 2\beta\left[m^2_{H_2} - m^2_{H_1} - 2(m^2_{\tau^c} + m^2_L)\right]\cr
&=0.49\left(\frakk{m_0}{40}\right)^2\TeV^2.\cr}}
(The numerical results above apply for $\tan\beta=5$.)
We have demonstrated that it is possible to construct a viable  model by
combining the AMSB scenario with FI $D$-terms in a model with  an extra
$U_1$.  The model incorporates natural flavour conservation 
and suppression of proton decay. 
One might imagine a more elegant version of the model  which
forbade the $\mu_{s}$-term, and incorporated neutrino masses; this is
not possible, however, without  introducing fields which are MSSM
non-singlets\cham. A recent version  of this idea (not in the AMSB
context)  is to be found in Ref.~\ref\aoki{M.~Aoki and N.~Oshimo,
hep-ph/9907481}; however because in this case $SU_3$ is not
asymptotically  free due to the presence of extra colour triplets, it is
hard  (in the AMSB framework) to achieve an acceptable vacuum.  

\bigskip\centerline{{\bf Acknowledgements}}\nobreak

This work was supported in part by the 
Leverhulme Trust. We thank Nigel Backhouse, Victor Flynn  and Roz 
Wild for Diophantine discussions. 

\listrefs

\end